\documentclass[aps,prl,twocolumn,showpacs,showkey,square,numbers,amssymb,amsmath,nobibnotes,footinbib]{revtex4}
\usepackage{bm}
\usepackage{amsmath}
\usepackage{amsfonts}
\usepackage{amssymb}   
\usepackage{eucal}
\usepackage{verbatim}
\usepackage{times}
\usepackage{graphicx}

\def\Ord{\mathcal{O}}
\newcommand{\barr}{\begin{eqnarray}}
\newcommand{\earr}{\end{eqnarray}}

\begin{document}

% Use the \preprint command to place your local institutional report
% number in the upper righthand corner of the title page in preprint mode.
% Multiple \preprint commands are allowed.
% Use the 'preprintnumbers' class option to override journal defaults
% to display numbers if necessary
%\preprint{}

%Title of paper
\title{Number of relevant directions in Principal Component Analysis and Wishart random matrices.}

% repeat the \author .. \affiliation  etc. as needed
% \email, \thanks, \homepage, \altaffiliation all apply to the current
% author. Explanatory text should go in the []'s, actual e-mail
% address or url should go in the {}'s for \email and \homepage.
% Please use the appropriate macro foreach each type of information

% \affiliation command applies to all authors since the last
% \affiliation command. The \affiliation command should follow the
% other information
% \affiliation can be followed by \email, \homepage, \thanks as well.

\author{Satya N. Majumdar and Pierpaolo Vivo}
\affiliation{Laboratoire de Physique Th\'{e}orique et Mod\`{e}les
Statistiques (UMR 8626 du CNRS), Universit\'{e} Paris-Sud,
B\^{a}timent 100, 91405 Orsay Cedex, France}

\date{\today}

\begin{abstract}

We compute analytically, for large $N$, the probability $\mathcal{P}(N_+,N)$ that 
a $N\times N$ Wishart random matrix has $N_+$ eigenvalues exceeding a threshold 
$N\zeta$, including its large deviation tails. This probability plays a benchmark 
role when performing the Principal Component Analysis of a large empirical 
dataset. We find that $\mathcal{P}(N_+,N)\approx\exp(-\beta N^2 \psi_\zeta(N_+/N))$, 
where $\beta$ is the Dyson index of the ensemble and $\psi_\zeta(\kappa)$ is a 
rate function that we compute explicitly in the full range $0\leq \kappa\leq 1$ 
and for any $\zeta$. The rate function $\psi_\zeta(\kappa)$ displays a quadratic 
behavior modulated by a logarithmic singularity close to its minimum 
$\kappa^\star(\zeta)$. This is shown to be a consequence of a phase transition in 
an associated Coulomb gas problem. The variance $\Delta(N)$ of the number of 
relevant components is also shown to grow universally (independent of $\zeta)$ 
as 
$\Delta(N)\sim (\beta 
\pi^2)^{-1}\ln N$ for large $N$.

\end{abstract}

% insert suggested PACS numbers in braces on next line
% insert suggested keywords - APS authors don't need to do this

%\maketitle must follow title, authors, abstract, \pacs, and \keywords

\maketitle

%\textit{Introduction -} 
Large datasets, such as financial or climate data, typically contain 
a mixture of relevant fluctuations responsible for hidden inherent patterns
and irrelevant minor fluctuations. Compressing
the data by filtering out the irrelevant fluctuations, while retaining
the relevant ones, is crucial for many practical applications. The most 
commonly used technique for such data compression is the "Principal
Component Analysis" (PCA) with an impressive list of applications
including image
processing~\cite{Wilks,Fukunaga,Smith}, biological
microarrays~\cite{arrays1,arrays2}, population
genetics~\cite{Cavalli,genetics}, finance~\cite{BP,Burda}, meteorology
and oceanography~\cite{Preisendorfer}. The basic idea of PCA is
rather simple. Consider for instance an experiment where 
a set of $N$ observables is recorded $M$ times. The collected data can 
be arranged into a rectangular $M\times N$ matrix $\mathbf{X}$ and
adjusted to have zero mean. 
For example, $(\mathbf{X})_{ij}$ might
represent examination marks of the $i$-th student ($1\le i\le M$) in
the $j$-th subject (physics, mathematics, chemistry, etc.), or 
the recorded temperatures
of the $i$-th largest American city on the $j$-th measurement day.
The product symmetric $(N\times N)$ square matrix $\mathbf{C}=\mathbf{X}^T 
\mathbf{X}$
represents the empirical covariance (unnormalized) matrix of the data that
encodes all correlations.   

In PCA, one collects
eigenvalues and eigenvectors of $\mathbf{C}$. 
The eigenvector (\emph{principal component})
associated with the largest eigenvalue $\lambda_1$ gives the 
direction along which the data are maximally scattered and correlated.
The scatter and correlations progressively reduce as one considers
lower and lower eigenvalues. Next
one retains only the top $N_+$ eigenvalues (out of $N$) and their corresponding
eigenvectors, while discarding all the lower eigenvalues and eigenvectors.
Subsequently, the data can be re-expressed in a 
new basis with the following desirable features (i) the dimension of the new 
basis is \emph{smaller}
(i.e. redundant information has been wiped out), and (ii) the 
total variance of data captured along the new eigendirections 
is \emph{larger} (i.e. only the most important
patterns present in the initial data have been retained).

While this procedure is well defined in principle, one immediately faces a 
practical problem, namely: what is the optimal number of eigenvalues $N_+$ to 
retain in a given data set? 
Unfortunately, a sound prescription for the choice of $N_+$ is generally lacking, 
and one has to resort to 
any of the \emph{ad hoc} 
stopping criteria available in the literature \cite{jack}.
For instance, in the commonly used Kaiser-Guttman type rule \cite{guttman}
one keeps the eigenvalues greater than a threshold 
value $\zeta N$, where the threshold may represent, e.g, the
average variance of the empirical dataset. 

However, this simple operational rule has attracted some criticism \cite{critic}
since randomly generated data (with no underlying pattern whatsoever) 
may also produce eigenvalues exceeding any empirical threshold, just by 
statistical fluctuations. It is thus compelling to 
define a statistical criterion able to discriminate whether the observed 
number of eigenvalues $N_+$ exceeding a chosen threshold $\zeta N$ retain
inherent patterns of the data or just reflect random noise. 

A Bayesian
approach provides a possible answer to this problem \cite{bapp}.
Suppose we observe a number $N_+$ of eigenvalues of $\mathbf{C}$ exceeding a 
fixed threshold decided by some stopping rule. We need to 
estimate the probability $P(M_i|N_+)$
that the data are extracted from a certain model $M_i$ 
(out of a set of possible $\{M_j\}$), \emph{given} the observed event $N_+$. 
According to Bayes' rule:
\begin{equation}
P(M_i|N_+)=
\frac{P(N_+|M_i)P(M_i)}{\sum_j P(N_+|M_j)
P(M_j)}\label{bayes}
\end{equation}
where $P(M_i)$ is the \emph{prior}, i.e. the probability 
that the data come from model $M_i$ without the knowledge
of the observed event $N_+$, while 
$P(N_+|M_i)$
is called the \emph{likelihood} of drawing the value $N_+$ if the data are taken 
from the model $M_i$. Assuming that one has some apriori knowledge of the
priors, the most crucial ingredient in Bayesian analysis is the estimate
of the likelihood $P(N_+|M_i)$ for a model $M_i$.  
The most natural `null' model to compare data with, in our context, is the 
random model where the 'data' $X_{ij}$
are replaced by pure noise, say from a Gaussian distribution with zero mean
and unit variance. In that case, the corresponding covariance matrix
$\mathbf{W}=\mathbf{C}= \mathbf{X^\dagger X}$ is called the Wishart 
matrix~\cite{Wishart}.  
Thus, the natural and important question that one faces is: what is the 
likelihood 
$P(N_+|\mathbf{W})$ of $N_+$ associated with the Wishart matrix model?
 
In this Letter, by suitably adapting a Coulomb gas method recently 
used~\cite{vivoindex1,vivoindex2}
to compute the distribution of the number of positive eigenvalues
of Gaussian random matrices, we are able to compute analytically
this likelihood $P(N_+|\mathbf{W})$ for arbitrary $\zeta$ and large $N$. To make 
the $N$ dependence explicit,
henceforth we use the notation $P(N_+|\mathbf{W})\equiv {\mathcal 
P}(N_+,N)$ where ${\mathcal
P}(N_+,N)$ then denotes the full probability distribution
of the number of eigenvalues of an $(N\times N)$ Wishart matrix exceeding a fixed 
threshold $\zeta N$. In addition
to serving as a crucial benchmark for the 
Bayesian analysis of the number of retained components
in PCA, we show that the distribution ${\mathcal
P}(N_+,N)$ also has a very rich and beautiful 
structure. 

\begin{figure}[htb]
\begin{center}
\includegraphics[bb =0 0 628 406,totalheight=0.23\textheight]{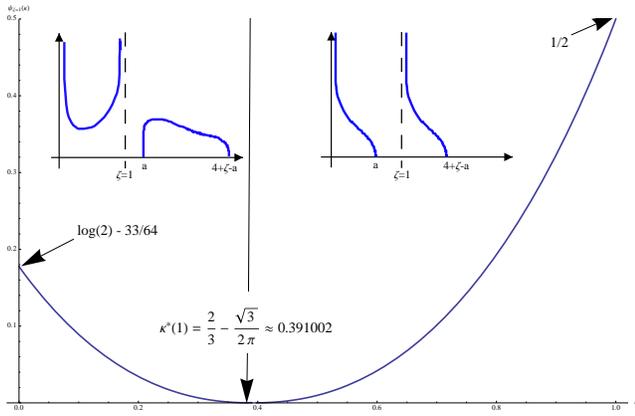}
\caption{The rate function $\psi_\zeta$ as a function of $\kappa\in [0,1]$ from 
Eq. \eqref{psifinal} for $\zeta=1$. In the two insets, a sketch of the 
corresponding Coulomb gas equilibrium densities
(Eq. \eqref{solutionforf}) in the two regimes $\kappa<\kappa^\star(1)$ (left) and 
$\kappa>\kappa^\star(1)$ (right). Only at $\kappa=\kappa^\star(1)$ the two 
disconnected blobs of particles on 
both sides of the barrier at $\zeta=1$ merge into a 
single continuous density (the usual Mar\v cenko-Pastur 
distribution $f_{\mathrm{mp}}(x)$).}
\label{psikappa}
\end{center}
\end{figure}

Let us first summarize our main results. We consider
Wishart matrices $\mathbf{W}=\mathbf{X^\dagger X}$ where $\mathbf{X}$
is in general a rectangular $M\times N$ matrix ($M\geq N$) 
with independent entries (real, complex or quaternions, 
labelled by the Dyson index $\beta=1,2,4$) drawn from a standard 
Gaussian distribution of zero mean and unit variance. 
We focus, for convenience, on the case when $M-N\sim \mathcal{O}(1)$. 
The matrix $\mathbf{W}$ has $N$ non-negative random eigenvalues.
We
compute the distribution ${\mathcal
P}(N_+,N)$ for large $N$ where $N_+$ is the number
of eigenvalues of $\mathbf{W}$ exceeding the threshold $\zeta N$.
Setting $N_+=\kappa N$, we show that ${\mathcal
P}(N_+=\kappa N, N)$ behaves for large $N$ as~\cite{remark1}
\begin{equation}
\mathcal{P}(N_+=\kappa N,N)\approx
\exp\left[-\beta N^2 \psi_\zeta(\kappa)\right],
\quad 0\leq \kappa\leq 1\label{general1}
\end{equation}
where the rate function $\psi_\zeta(\kappa)$ (independent of $\beta$) 
is computed
analytically for arbitrary $\zeta$ (see Eq. \eqref{psifinal}) and
plotted in Fig. \ref{psikappa} for $\zeta=1$. The rate function
has a minimum (zero) at the critical value $\kappa= \kappa^\star(\zeta)$
where $\kappa^\star(\zeta)=(4\ \mathrm{arcsec}(2/\sqrt{\zeta}) -
\sqrt{\zeta (4 - \zeta)})/2 \pi$. Thus the distribution
is peaked around $N_+=\kappa^\star(\zeta) N$ which
is precisely its mean value $\langle N_+\rangle= \kappa^\star(\zeta) N$.
Around this critical value, the rate function is 
non-analytic and displays a quadratic behavior modulated by a
universal ($\zeta$-independent) logarithmic
singularity
\begin{equation}
\psi_\zeta(\kappa^\star(\zeta)+\delta)\sim -(\pi^2/2)\delta^2/\log |\delta|\, .
\label{logsingular}
\end{equation}
The physical origin of this non-analytic behavior is traced back to a phase 
transition in the
associated Coulomb gas problem when $\kappa$ crosses the critical value
$\kappa^\star(\zeta)$. Inserting this expression in \eqref{general1},
one finds that, for small fluctuations of $N_+$ around its mean $\langle N_+\rangle$ 
on a scale $\sqrt{\ln N}$, the
distribution has a Gaussian form, ${\mathcal
P}(N_+,N) \sim \exp\left[-\left(N_+ -\kappa^\star(\zeta)\,N\right)^2/{2\Delta( 
N)}\right]$ where the variance 
\begin{equation}\label{varianceN}
\Delta(N)=N^2\langle (\kappa-\kappa^\star(\zeta))^2\rangle
\sim\frac{1}{\beta \pi^2}\ln N+\Ord(1)
\end{equation}
This leading behavior is also universal, i.e., independent of $\zeta$ and
is the same as the variance of the number of positive 
eigenvalues
in the Gaussian ensembles \cite{vivoindex1,vivoindex2}. 
We thus conclude that on a scale 
$\sim\mathcal{O}(\sqrt{\ln N})$, the distribution $\mathcal{P}(N_+,N)$ is 
Gaussian with mean $N_+=\kappa^\star(\zeta)$ and variance in \eqref{varianceN}, 
but with non-Gaussian
large deviation tails that are described by the general 
formulae \eqref{general1} and \eqref{psifinal}. 

\begin{figure}[htb]
\begin{center}
\includegraphics[bb =   -260   173   874   618,
totalheight=0.15\textheight]{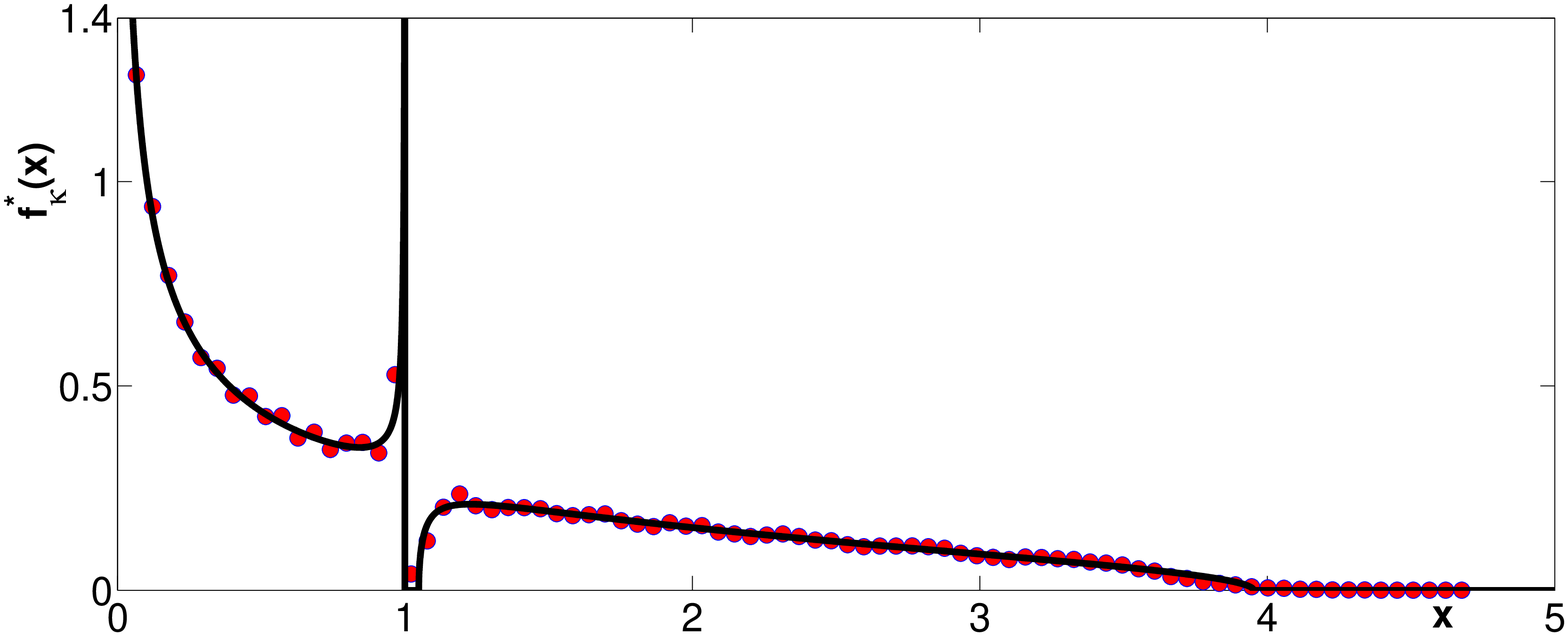}\\
%\vspace{5pt}
\includegraphics[bb = -260   173   874   619,totalheight=0.15\textheight]{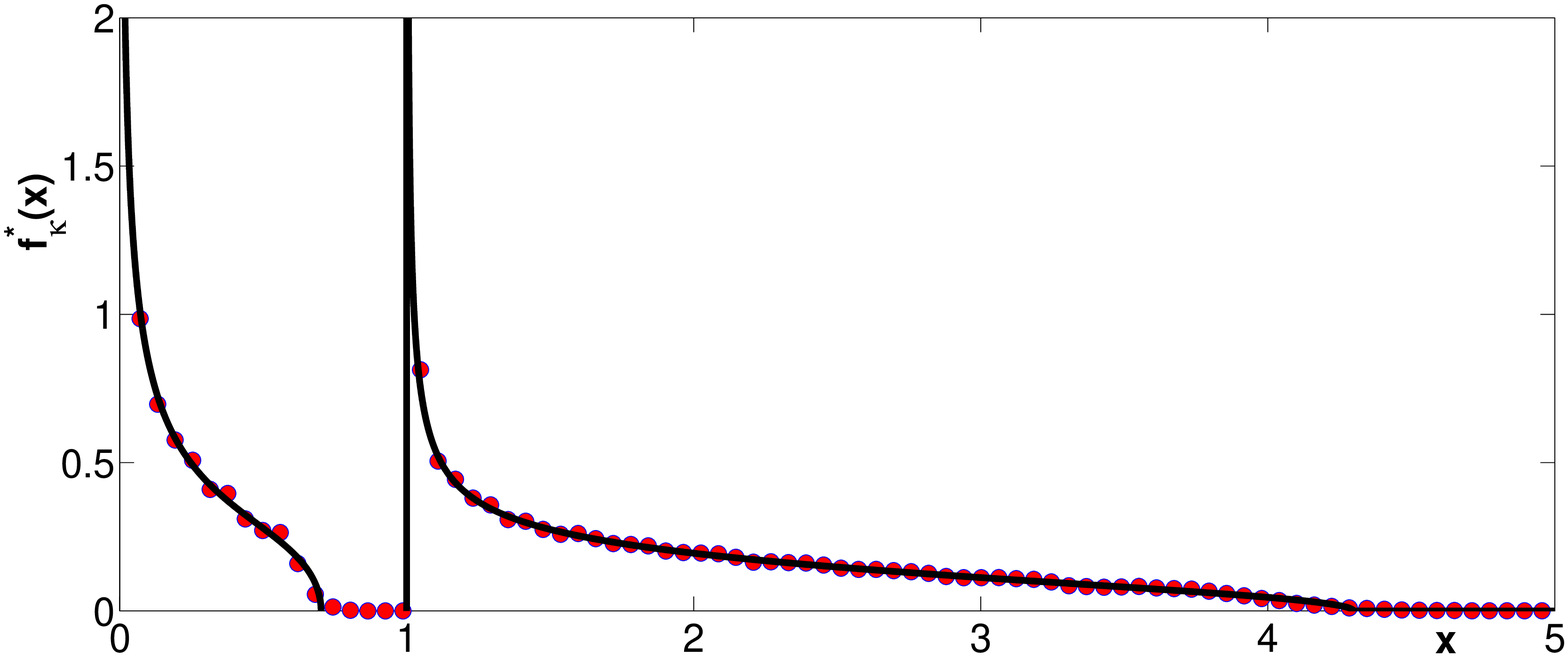}
\caption{Equilibrium density of the Coulomb fluid, Eq. \eqref{solutionforf} 
for $\zeta=1$ (black solid line). The density in general has a disconnected 
support and exists in two different phases, depending
on whether $\kappa$ is smaller (top panel) or larger (bottom panel) 
than $\kappa^\star(\zeta)$. Red dots correspond to Montecarlo simulations of a 
Coulomb gas of $20$ charges, where $7$ charges (top)
and $12$ charges (bottom) were constrained to the right of 
$\zeta=1$ \cite{supp}. The corresponding values of the edge point 
$a$ are $a=1.04815>\zeta$ (top) and $a=0.701721<\zeta$ (bottom). The area under the 
leftmost blob is $1-\kappa$ and under the rightmost blob is $\kappa$.}
\label{densityfigs}
\end{center}
\end{figure}

We start by recalling some well known spectral properties of Wishart matrices
(also known as Laguerre or chiral ensemble) defined earlier.
The $N$ non-negative eigenvalues $\{\lambda_i \}$’of the
Wishart matrix ${\mathbf W}$ are distributed via the joint
probability density law~\cite{James}
\begin{equation}
\mathcal{P}(\lambda_1,\ldots,\lambda_N)=\frac{1}{Z_N}
e^{-\frac{\beta}{2}\sum_{i=1}^N\lambda_i}\prod_{i=1}^N
\lambda_i^{\frac{\beta}{2}\alpha}
|\Delta(\mbox{\boldmath$\lambda$})|^\beta\label{jpdf}
\end{equation}
where $\Delta(\mbox{\boldmath$\lambda$})=\prod_{j<k}(\lambda_j-\lambda_k)$, 
$\alpha=(1+M-N)-2/\beta$ and $Z_N$ is the normalization constant.
It is also known that for large $N$, $Z_N\simeq \exp[-\beta \Omega_0 N^2]$
where $\Omega_0=3/4$~\cite{Mehta,vivo2007large}. 
This joint law \eqref{jpdf} can be conveniently recast
in the Boltzmann form
\begin{equation}
\mathcal{P}(\lambda_1,\ldots,\lambda_N)\propto\exp(-(\beta/2) 
E[ \{\mbox{\boldmath$\lambda$}\}])\label{jpddyson}
\end{equation}
where the energy of a configuration $\{\mbox{\boldmath$\lambda$}\}$ is
\begin{equation}
E[ \{\mbox{\boldmath$\lambda$}\}]=
\sum_{i=1}^N \lambda_i-\alpha\sum_{i=1}^N 
\ln\lambda_i-\sum_{j\neq k}\ln |\lambda_j-\lambda_k|\label{en}
\end{equation}
Here, $\beta/2$ stands for the inverse temperature and for simplicity we 
focus here on almost-square matrices with $M-N$ (and thus $\alpha$) of 
$\mathcal{O}(1)$ for large $N$.
This thermodynamical analogy, originally due to Dyson \cite{Dyson}, allows to treat 
the system of eigenvalues as a Coulomb gas, i.e., a fluid of charged particles
confined to the positive half-line by two competing interactions: the 
external linear + logarithmic  
potential (the first two terms) in \eqref{en} tends to push the charges towards 
the origin, while the third term representing mutual logarithmic repulsion 
between any pair of charges
tends to spread them apart. It is useful first to estimate the
typical scale of an eigenvalue $\lambda$ for large $N$. The
first two terms typically scale for large $N$ as $\lambda_{\rm typ} N$
and $\ln (\lambda_{\rm typ})\, N$. In contrast, the pairwise Coulomb repulsion
(the third term) typically scales as $N^2$ for large $N$. Balancing the
first and the third term indicates $\lambda_{\rm typ}\sim N$ for large $N$.
Consequently, as long as $\alpha\sim \mathcal{O}(1)$, the second term becomes smaller 
($\sim \mathcal{O}(N)$) compared to the other
two terms ($\sim \mathcal{O}(N^2)$) and hence can be neglected.
One then expects the average spectral density (normalized to unity)
$\rho(\lambda)={N^{-1}}\sum_i \langle \delta(\lambda-\lambda_i)\rangle $ 
to have the scaling form for large $N$:  
$\rho(\lambda)=N^{-1}\,f_{\mathrm{mp}}(\lambda/N)$. The scaling
function $f_{\mathrm{mp}}(x)=(2\pi)^{-1}\sqrt{(4-x)/x}$ is the celebrated Mar\v 
cenko-Pastur (MP)
distribution on the compact support $x\in (0,4)$~\cite{MP}.

Given the joint distribution in \eqref{jpdf}, our goal is to compute
the statistics of $N_+$ denoting the number of eigenvalues greater
than a fixed threshold value $\zeta N$, i.e., $N_+= \sum_{i=1}^N 
\theta(\lambda_i- \zeta N)$ where $\theta(x)$ is the Heaviside step function.
The average value of $N_+$ can be easily estimated from the MP spectral density,
$\langle N_+\rangle= N\, \int_{\zeta}^4 f_{\mathrm{mp}}(x)dx= \kappa^\star (\zeta) N$
where $\kappa^*(\zeta)= (4\ \mathrm{arcsec}(2/\sqrt{\zeta}) -
\sqrt{\zeta (4 - \zeta)})/2 \pi$. The full probability 
distribution of $N_+$ can be written as the $N$-fold integral  
\begin{equation}\label{probdef}
\mathcal{P}(N_+,N)\propto\int\prod_{i=1}^N d\lambda_i 
e^{-\beta E[ \{\mbox{\boldmath$\lambda$}\}]}
\delta\left(N_+ -\sum_{i=1}^N \theta(\lambda_i-\zeta N)\right)
\end{equation}
which we evaluate next for large $N$ by extending the Coulomb
gas analogy mentioned above.  \\
%\vspace{3pt}

The evaluation of the $N$-fold integral \eqref{probdef} in the large $N$ limit 
consists of the following steps: first, we introduce an integral representation 
for the $\delta$ function, $\delta(x)=(2\pi)^{-1}(\beta N/2)\int dp 
\exp(\mathrm{i}\beta Npx/2)$. Then, we cast again the integrand in the Boltzmann 
form $\exp\left[-(\beta/2) E_\kappa(\{\mbox{\boldmath$\lambda$}\})\right]$ with 
$E_\kappa(\{\mbox{\boldmath$\lambda$}\}) = E(\{\mbox{\boldmath$\lambda$}\}) +
 A_1 N \left(\sum_{i}\theta(\lambda_i-\zeta N)-\kappa N\right)$,
where $A_1=\mathrm{i}p$ can be interpreted as a Lagrange multiplier.
Written in this form, the integral has again a natural thermodynamical 
interpretation as the canonical partition function of a Coulomb gas in 
equilibrium at inverse temperature $\beta/2$. This time, however, in addition to 
the linear confinement and the logarithmic repulsion, the fluid particles 
(eigenvalues) are subjected to another (discontinuous) external 
potential. This 
external term involving $A_1$ in the new energy function 
$E_\kappa(\{\mbox{\boldmath$\lambda$}\})$ has the effect of constraining a 
fraction $\kappa$ of the fluid charges to the right of the point $x=\zeta N$.

In the large $N$ limit, the Coulomb gas with $N$ discrete charges becomes
a continuous gas which can be described by a continuum 
(normalized to unity) density 
function $\varrho(\lambda)=N^{-1}\sum_{i=1}^N\delta(\lambda-\lambda_i)$.
Consequently, one can replace the original multidimensional integral in
\eqref{probdef} by a functional integral over the space of 
 $\varrho(\lambda)$. This procedure,
originally introduced by Dyson~\cite{Dyson}, has 
recently been used successfully in a number of different contexts (see 
\cite{DM1,DM2,vivo2007large,vivoindex1,vivoindex2} and references therein).
Noting further that 
the density $\varrho(\lambda)$ is expected to have the scaling form 
$\varrho(\lambda)=N^{-1}f_{\kappa}(\lambda/N)$ for large $N$,
each term in the energy $E_\kappa(\{\mbox{\boldmath$\lambda$}\})$ is of the same order 
$\sim\mathcal{O}(N^2)$ and is expressed as an integral over the function 
$f_\kappa(x)$ (see eq. \eqref{action} below). The scaled density $f_{\kappa}(x)$ 
satisfies the overall normalization condition $\int_{0}^\infty dx f_{\kappa}(x) 
=1$ and the constraint $\int_{0}^\zeta dx f_{\kappa}(x) =1-\kappa$ arising from 
the theta-function term in $E_\kappa(\{\mbox{\boldmath$\lambda$}\})$.

The probability \eqref{probdef} can now be rewritten as 
$\mathcal{P}(N_+=\kappa N,N)=Z_\kappa(N)/Z_N$,
where the numerator $Z_\kappa(N)$ is the following functional integral over 
$f_\kappa(x)$, supplemented by two additional integrals 
over auxiliary variables $A_1$ and $A_2$ 
enforcing the two constraints mentioned above:
\begin{equation}
Z_\kappa(N) =\int d A_1\ d A_2\ \mathcal{D}[f_{\kappa}(x)]
\exp\left\{-\frac{\beta}{2}N^2 \mathcal{S}[f_\kappa(x)]\right\}.
\label{Zkappa}
\end{equation} 
The action $\mathcal{S}[f_\kappa(x)]$ is given by:
\begin{widetext}
\begin{equation}
\mathcal{S}[f_\kappa(x)]=
\int_{0}^\infty dx x f_{\kappa}(x)-\int_{0}^\infty\int_{0}^\infty 
dx dx^\prime f_{\kappa}(x) f_{\kappa}(x^\prime)\ln|x-x^\prime|+
A_1\left(\int_{0}^\zeta dx f_\kappa(x)-(1-\kappa)\right)+ 
A_2\left(\int_{0}^\infty dx f_\kappa(x)-1\right) \label{action}
\end{equation}
\end{widetext}

Eq. \eqref{Zkappa} is indeed amenable to a saddle point analysis:
\begin{equation}
\label{znum} Z_\kappa(N) \approx \exp\left(-\frac{\beta}{2}N^2 
\mathcal{S}[f_\kappa^\star(x)]\right),\quad  Z_N \approx \exp(-\beta\Omega_0 N^2)
\end{equation}
where $\Omega_0=3/4$ and $f_\kappa^\star(x)$ is obtained from  
$\frac{\delta \mathcal{S}[f_\kappa(x)]}{\delta f_\kappa}=0$. This
gives
\begin{equation}\label{saddle point}
x+A_1\theta(\zeta-x)+A_2=2\int dx^\prime f_\kappa^\star(x^\prime)\log|x-x^\prime|
\end{equation} 
for all $x\in\mathsf{supp}(f_\kappa^\star(x))$ where
$\mathsf{supp}$ denotes the region when charges exist, i.e., $f_{\kappa}^\star(x)$
is positive.
Varying the action with respect 
to 
$A_1$ and $A_2$ just reproduces the two constraints. The function 
$f_\kappa^\star(x)$ obtained in this way is 
the rescaled equilibrium density of a Coulomb fluid where a fraction $\kappa$
of particles are constrained to lie to the right of a barrier at $x=\zeta$. 
It is easy to show~\cite{supp} that $f_{\kappa}^\star(x)$ reduces to the unconstrained 
MP density 
$f_{\mathrm{mp}}(x)$ when $\kappa=\kappa^\star(\zeta)$.

Thus, for large $N$, the saddle point analysis precisely predicts
the result in \eqref{general1} with the rate function given by
\begin{equation}
\psi_\zeta(\kappa)= \frac{1}{2}\mathcal{S}[f_\kappa^\star(x)]-\Omega_0 \,.
\label{largedevlaw}
\end{equation}
%\begin{equation}
%\mathcal{P}(N_+=\kappa N,N)\approx\exp\left[-\beta N^2 
%\underbrace{\left(\frac{1}{2}\mathcal{S}[f_\kappa^\star(x)]-
%\Omega_0\right)}_{\psi_\zeta(\kappa)}\right].
%\label{largedevlaw}
%\end{equation}
It then remains to solve the saddle-point equation \eqref{saddle point}. 
Taking one more derivative with respect to $x$ ($x\neq 0$):
\begin{equation}
\frac{1}{2} = \mathrm{Pr}\int \frac{f_\kappa^\star(x^\prime)}{x-x^\prime}
dx^\prime\label{tricomi}
\end{equation}
where $\mathrm{Pr}$ denotes Cauchy's principal value, supplemented with the 
constraints $\int_{0}^\infty dx f_{\kappa}^\star(x)=1$ and $\int_{0}^\zeta dx 
f_{\kappa}^\star(x) =1-\kappa$.

To invert the singular integral equation \eqref{tricomi} is a nontrivial
challenge. One often needs to guess the solution and verify it a posteriori.
To get a feeling how the solution may look like, we first did a Monte Carlo
simulation of the Coulomb gas which brought out the following very
interesting features.
For $\kappa\neq \kappa^\star(\zeta),0$ and $1$, the equilibrium charge density 
$f_\kappa^\star (x)$ 
(i) generally consists 
of two disconnected supports: a blob of $(1-\kappa)N$ eigenvalues to the left of 
$\zeta$ separated from a second blob of $\kappa N$ eigenvalues to the right of 
$\zeta$ (see Fig. \ref{densityfigs}), and (ii) the actual shapes of the 
two blobs depend on whether $\kappa<\kappa^\star(\zeta)$ (top panel of Fig. 
\ref{densityfigs}) or $\kappa>\kappa^\star(\zeta)$ (bottom panel of Fig. 
\ref{densityfigs}), i.e. the fluid undergoes a phase transition as $\kappa$ is 
varied across the critical point $\kappa^\star(\zeta)$. When $\kappa\to 
\kappa^\star(\zeta)$, the two blobs merge into a single support solution
which is precisely the MP spectral density $f_{\mathrm{mp}}(x)$.

Extracting this two-support solution for a generic 
$\kappa\neq\kappa^\star(\zeta),0, 1$ from the singular 
integral equation \eqref{tricomi} poses the main technical challenge
that we have succeeded in solving. There exists a well known Riemann-Hilbert
method~\cite{scalar} for solving such singular integral equations when
the solution has a single support~\cite{Tricomi}. For two disjoint supports,
this method~\cite{scalar} was recently generalized in a number
of different problems~\cite{vivoindex1,vivoindex2,cond1,cond2,cond3}.
Adapting this generalized method to our problem (for details
see \cite{supp}), we find that the two-support solution of \eqref{tricomi}
satisfying the two constraints is given explicitly by the expression
\begin{equation}
f_\kappa^\star(x)=
\frac{1}{2\pi}\sqrt{\frac{(x-a)(4+\zeta-a-x)}{x(x-\zeta)}}. \label{solutionforf}
\end{equation}
The edge point $a$, depending on $\kappa$, is to be determined as 
the unique solution of the constraint $\int_{0}^\zeta dx f_{\kappa}^\star(x) =
1-\kappa$,
which can be rewritten in terms of generalized hypergeometric functions \cite{supp}. 
We have $a>\zeta$ for $\kappa<\kappa^\star(\zeta)$, $a<\zeta$ for 
$\kappa>\kappa^\star(\zeta)$
and $a=\zeta$ for $\kappa=\kappa^\star(\zeta)$. The density 
$f_\kappa^\star (x)$ is supported on the union of two disconnected 
intervals for $\kappa\neq \kappa^\star(\zeta)$, $[0,\zeta]\cup 
[a,4+\zeta-a]$ (if $\kappa\leq\kappa^\star(\zeta)$) or 
$[0,a]\cup [\zeta,4+\zeta-a]$ (if $\kappa\geq\kappa^\star(\zeta)$) [see, e.g.,
the insets of Fig. \ref{psikappa}].
For $\kappa=\kappa^\star(\zeta)$, the two intervals merge into a single one and 
$f_\kappa^\star (x)=f_{\mathrm{mp}}(x)$ as expected. In the extreme cases 
$\kappa=0,1$, the density has also a single support
and its analytical form is known from earlier works \cite{vivo2007large}. 
The function $f_\kappa^\star (x)$ is plotted in Fig. \ref{densityfigs} for
a specific choice $\zeta=1$, 
together with the result of Montecarlo simulations of the Coulomb gas,
showing excellent agreement. 

Next we substitute this explicit solution $f_{\kappa}^\star (x)$ in 
\eqref{action} to evaluate the 
saddle-point action $\mathcal{S}[f_\kappa^\star (x)]$. For this, we need to 
compute 
the 
single and double integrals in \eqref{action}.
The double integral can be written in terms of a simple integral after 
multiplying \eqref{saddle point} by $f_\kappa^\star(x)$ and 
integrating over $x$. This leads to:
\begin{equation}
\mathcal{S}[f_\kappa^\star(x)]= \frac{1}{2}\left[\mu_1(\zeta) - A_1 (1-\kappa) - 
A_2\right]
\label{action2}
\end{equation}
where $\mu_n(\zeta)= \int_{0}^{\infty} x^n f_\kappa^\star(x) dx$ is the 
$n$-th moment of the density. The first moment can be explicitly computed as 
$\mu_1(\zeta)=1-a+\frac{a^2}{4}+\zeta-\frac{a\zeta}{4}$. It remains to fix the 
two Lagrange multipliers $A_1$ and $A_2$, assigning suitable $x$ 
values in Eq. \eqref{saddle point}. Skipping details~\cite{supp}, it turns out 
that $A_1$
and $A_2$ can be expressed in terms of a pair of functions:
\begin{equation}
F_{(\pm)}=\frac{1}{2}\pm\frac{1}{2}\sqrt{\frac{(a-x)(4+\zeta-a-x)}{x(x-\zeta)}} 
\end{equation}
defined for $x \notin\mathsf{supp}(f_\kappa^\star (x))$. It can be shown that 
these functions are essentially the moment generating functions of 
$f_\kappa^\star (x)$ \cite{supp}.

Combining \eqref{action2} and \eqref{largedevlaw}, 
the final expression for the rate function $\psi_\zeta(\kappa)$ reads:
\begin{widetext}
\begin{equation}\label{psifinal}
\psi_\zeta(\kappa)=
\begin{cases}
\frac{1}{4}\left[\tilde\mu(\zeta)-  (1-\kappa) \left(a-\zeta-2\int_\zeta^a 
F_{(+)}(x)dx\right)- 
2\ln(4+\zeta-a)+2\int_{4+\zeta-a}^\infty dx\left(F_{(-)}(x)-\frac{1}{x}\right)\right],
&\kappa\leq\kappa^\star(\zeta)\\
\frac{1}{4}\left[\tilde\mu(\zeta)-  (1-\kappa) 
\left(\zeta-a-2\int_a^\zeta F_{(-)}(x)dx\right)- 
2\ln(4+\zeta-a)+2\int_{4+\zeta-a}^\infty dx\left(F_{(-)}(x)-\frac{1}{x}\right)\right],
&\kappa\geq\kappa^\star(\zeta)
\end{cases}
\end{equation}
\end{widetext}
where $\tilde\mu(\zeta)=2-2a+a^2/4+2\zeta-a\zeta/4$. 
The behavior of $\psi_\zeta(\kappa)$ around the minimum 
$\kappa=\kappa^\star(\zeta)$ in \eqref{logsingular}
is found after a 
lengthy but straightforward expansion \cite{supp}.

In summary, using a Coulomb gas approach we computed analytically the 
likelihood $P(N_+|\mathbf{W})$ 
of retaining $N_+$ principal 
components if a 
completely random (pure noise) $N\times N$ prior ${\mathbf W}$ (Wishart random 
matrix) 
is assumed for the underlying data. This likelihood is an essential ingredient in 
Bayes' formula 
\eqref{bayes}, which in turn gives the probability that a 
certain number $N_+$ of significant eigenvalues is observed
if the data represents pure noise. Thus the rate function computed in 
\eqref{psifinal} serves indeed 
as a benchmark to gauge the significance of results obtained for empirical data. 
We also found an interesting phase transition in the associated Coloumb
gas problem when two separated blobs of Coulomb charges merge onto a
single one as a critical value $N_+/N=\kappa=\kappa^\star(\zeta)$ is approached.
This phase transition is responsible for the 
appearance of a logarithmic singularity in the rate function 
close to its minimum, which in turn leads to a highly universal
logarithmic growth of the variance of $N_+$ with $N$.  

The present work can be developed in several directions: on one hand, it would be 
interesting to analyze the case $N\neq M$ and check if the universality found 
for $N\simeq M$ still persists. On the other hand, improved stopping criteria 
based on null random 
matrix results are very much called for. The Coloumb gas method developed
here for null random matrices may indeed be useful in that direction. 

S.N.M. acknowledges the support of ANR grant 2011-BS04-013-01 WALKMAT.

\end{document}